\documentclass[12pt,letterpaper]{article}
\pdfoutput=1
\usepackage{graphicx}
\usepackage{subfig}
\usepackage{amsfonts,amssymb}
\usepackage{amsthm}
\newcount{\myi}
\newcommand{\Repeat}[2]{%
    \myi=0
    \loop
        \ifnum\myi<#2
        #1
        \advance\myi by 1
    \repeat
}

\newcommand{\hrs}{hyperelliptic Riemann surface}
\newcommand{\hrss}{hyperelliptic Riemann surfaces}

\newcommand{\be}{\begin{equation}}
\newcommand{\ee}{\end{equation}}
\newcommand{\ben}{\begin{displaymath}}
\newcommand{\een}{\end{displaymath}}
\newcommand{\bea}{\begin{eqnarray}}
\newcommand{\eea}{\end{eqnarray}}
\newcommand{\bean}{\begin{eqnarray*}}
\newcommand{\eean}{\end{eqnarray*}}
\newcommand{\nn}{\nonumber \\}

\def\l {\lambda}
\def\a {\alpha}

\def\b {\beta}

\def\s {\sigma}
\def\e {\epsilon}

\renewcommand{\L}{\Lambda}
\renewcommand{\t}{\theta}

\newcommand{\calr}{\mbox{${\cal R}$}}

\newcommand{\commentout}[1]{}

\newcommand{\beq}{\begin{equation}}
\newcommand{\eeq}{\end{equation}}
\newcommand{\beqr}{\begin{displaymath}}
\newcommand{\eeqr}{\end{displaymath}}
\newcommand{\beqa}{\begin{eqnarray}}
\newcommand{\eeqa}{\end{eqnarray}}
\newcommand{\beqar}{\begin{eqnarray*}}
\newcommand{\eeqar}{\end{eqnarray*}}
\renewcommand{\k}{\kappa}
\newcommand{\m}{\mu}

\newcommand{\half}{\ensuremath{\frac{1}{2}}}

\newcommand{\adsn}[1]{$AdS_#1$ }

\newcommand{\hsp}{\hspace{0.2in}}
\newcommand{\vsp}{\vspace{2in}}

\begin{document}
%%%%%%%%%%%%%%%%%%%%%%%%%%%%%%%%%%%%%%%%%%%%%%%%%%%%%%%%%%%%%%%%%%%%%%%%
%%%%%%%%%%%%%%%%%%%%%%%%%%%%%%%%%%%%%%%%%%%%%%%%%%%%%%%%%%%%%%%%%%%%%%%%
%%%%%%%%%%%%%%%%%%%%%% TITLEPAGE %%%%%%%%%%%%%%%%%%%%%%%%%%%%%%%%%%%%%%%
%%%%%%%%%%%%%%%%%%%%%%%%%%%%%%%%%%%%%%%%%%%%%%%%%%%%%%%%%%%%%%%%%%%%%%%%
%%%%%%%%%%%%%%%%%%%%%%%%%%%%%%%%%%%%%%%%%%%%%%%%%%%%%%%%%%%%%%%%%%%%%%%%

\title{\LARGE \bf   Holographic Calculations of Euclidean  Wilson Loop Correlator in Euclidean anti-de Sitter Space}

\author{
%Martin Kruczenski\thanks{E-mail: \texttt{markru@purdue.edu}}\\
%        Department of Physics, Purdue University,  \\
%        525 Northwestern Avenue, W. Lafayette, IN 47907-2036,  \vspace{0.5cm} \\
%        {\em and} \vspace{0.5cm}\\
         Sannah Ziama\thanks{E-mail: \texttt{spziama@pa.uky.edu}}\\
                 Department of Physics and Astronomy, University of Kentucky,  \\
                505 Rose Street, Lexington, KY 40506-0055 }

\maketitle

\begin{abstract}

The correlation functions of  two or more Euclidean  Wilson loops of various shapes in Euclidean anti-de Sitter  space are computed  by considering the minimal area surfaces connecting the loops. The surfaces are parametrized  by Riemann theta functions associated with genus three \hrss.  In the case of two loops, the distance $L$ by which they are separated can be adjusted by continuously varying a  specific branch point of the auxiliary Riemann surface. When $L$ is much larger than the characteristic size of the loops, then the loops are approximately regarded as local operators and their correlator as the correlator of two local operators.   Similarly,  when a loop is very small compared to the size of  another loop, the small loop is considered as a local operator corresponding to a  light supergravity mode.  

\end{abstract}

\clearpage
\newpage

%\keywords{Classical string solutions, AdS/CFT, Wilson loops}

%\preprint{\tt{} \\
%          \tt{hep-th/yymmnnn}  }

%%%% INTRODUCTION

\section{Introduction}
Since the inception of the AdS/CFT correspondence \cite{malda}, many aspects of gauge theory have been made clearer by studying relevant quantities in string theory. One important example is the case of the Wilson loop, which is one of the most fundamental quantities of a gauge theory.  According to \cite{MRY,DGO}, a Wilson loop can be computed in  string theory by considering a minimal area surface in  anti-de Sitter space that ends on the loop in the boundary of the space.

 Many examples of Wilson loops have been studied in this context  \cite{cWL}, with much more general examples recently computed in  \cite{IKZ, KZ, K} where  a new, infinite parameter family of minimal area surfaces were studied. In those works the minimal area surfaces were constructed analytically in terms of Riemann theta functions associated to  hyperelliptic Riemann surfaces closely following previous work by  \cite{BB,BBook}.
 
Our objective in this paper is to holographically compute the correlation function of multiple Wilson loops of various shapes by means of minimal area surfaces. The AdS/CFT correspondence implies that the minimal area surfaces  have these loops as boundary. For two correlated Wilson loops, when one Wilson  loop is ``very small'' we will associate it with a particle emitted by the worldsheet  ending on the other loop \cite{BCFM}. Then the correlator in this case may by viewed as the correlator of a Wilson loop and a chiral primary operator.  Most of the technique used in this paper was established in \cite{IKZ, KZ} and the reader is encouraged to review them. 

 In a conformal field theory, it is often very useful to know the correlation functions of  nonlocal operators like the Wilson loop. In the limit when the t'Hooft coupling $\lambda$ is small the Wilson loop may be perturbatively expanded. For large $\lambda$, however, one may compute OPE of a Wilson loop correlator either by taking the correlator of the  loop with the relevant  chiral primary operator or by taking the correlator of the  loop with another loop \cite{BCFM}.

For example, the correlation of two circular Wilson loops of equal size $a$ separated by a distance $L$ has been well studied where, interestingly, it is shown that  a phase transition occurs when $L$ is large enough \cite{GO1, Zarem, Nian, BZ}. For small L the surface connecting the two loops is the catenoid. The transition occurs when L gets large enough to exceed a critical value when the minimal area surface prefers to separate into two hemispheres connected by a vanishingly narrow tube. 
 
The connected part of the correlators of two Wilson loops in the supergravity limit is 
 
 \beq
 \langle W^{\dagger}(C_{1} )W(C_{2}) \rangle_{conn} = \exp(-\frac{S}{2\pi \alpha'})
 \label{WLcorrelator}
 \eeq
 
 where S is  the Nambu-Goto action 
 \beq
 S = \int d\sigma d\tau \sqrt{\det  g_{\alpha \beta}} \,\, .
 \eeq
  Additionally, others \cite{ABT, AEKMS, AHS} have also studied  the correlator of null Wilson loops.

 The paper is organized as follows: In section two we review how to go from the sigma model to a nonlinear scalar equation of motion. Section three contains the main technical difference with the previous solutions \cite{KZ}. We set up the auxiliary Riemann surface in section four. Then we use the Riemann surface in section five to  compute  examples with two  loops.  In the following section we show how it may be extended to more than two Wilson loops, then we make concluding remarks in the final section.

%%%%%%%%%%%%%%%%%%%%%%%%%%%%%%%%%%%%%%%%%%%%%%%%%%%%%%%%%%%%%%%%%%%%%%%%%%%%%%%%%%%%%%%%%%%%%%%%%%%%%%%%%%%%%%%%%%%%

\section{Sigma Model in Euclidean \adsn3}

A convenient way to imagine Euclidean \adsn3 is to consider it as a  subspace of $\calr^{3,1}$ defined by the equation
\beq
  -X_0^2 + X_1^2 + X_2^2 + X_3^2 = -1\ ,
\label{Xsq1}
\eeq
with an obvious  $SO(3,1) \cong SL(2,\mathbb{C})$ global invariance. The action of the string in  conformal coordinates is given by 
\beqa
 S &=& \half \int \left( \partial X_\mu \bar{\partial} X^\mu - \Lambda (X_\mu X^\mu -1)\right) \ d\sigma\, d\tau \nn
    &=& \half \int \frac{1}{Z^2}\left(\partial_a X \partial^a X +\partial_a Y \partial^a Y +\partial_a Z \partial^a Z \right) \ d\sigma\, d\tau,
   \label{action}
\eeqa
where $\L$ is a Lagrange multiplier and the $\mu$ indices are raised and lowered with the $\calr^{3,1}$ metric
\beq
(v,w)=\sum_{i=1}^3{v_iw^i}-v_0w^0\,. 
\eeq
The complex coordinates $z$ and $\bar{z}$ are related to the world sheet coordinates, $\sigma_a=(\sigma,\tau)$ by  $z=\sigma+i\tau$, and $\bar{z}=\sigma-i\tau$. $X,Y,Z$ are called Poincar\'{e} coordinates (see (\ref{PoinX})). The string equations of motion are given by 
\beq
  \partial \bar{\partial} X_\mu = \Lambda X_\mu\ ,
\label{eomX}
\eeq
where $\L$, the Lagrange multiplier is given by 
\beq
\L = \partial X_\mu \bar{\partial} X^{\mu}=(X_z,X_{\bar{z}}). \label{Lcomp}
\eeq
The conformal condition is encoded in the Virasoro constraint
\beq
(X_z,X_z) = 0 =(X_{\bar{z}},X_{\bar{z}}).
\label{Vc1}
\eeq

The first step to solving the equations of motion is  to reduce the problem to an equation with a single unknown scalar field. In Euclidean \adsn3 space this scalar equation is the cosh-gordon equation
\beq
\partial\bar{\partial}\alpha=4 \cosh \alpha\,.
\label{chgordon}
\eeq
This reduction mechanism referred to as {\em Pohlmeyer reduction} \cite{Pohlmeyer} has been used in the study of many related problems. In the context of Minkowski space-time this procedure was used by Jevicki and Jin \cite{JJ} and by Kruczenski \cite{spiky}  to find new spiky string solutions, by Irrgang and Kruczenski \cite{IrK} to study new open string solutions in $AdS_{3}$, and by Alday and Maldacena \cite{AM2} to compute certain light-like Wilson loops. In a more geometric guise it was  employed  \cite{BB} to study constant mean curvature surfaces in hyperbolic space. We  review the idea and follow closely what was done  there.

In {\em Euclidean} $AdS_{3}$, form a basis 
\beq
t=(X,X_z,X_{\bar{z}},N)
\eeq
where 
\beq
(X,N)=(X_z,N)=(X_{\bar{z}},N)=0, (N,N)=1
\eeq
Note that (\ref{Xsq1}) and (\ref{Vc1}) imply
\beq
(X,X_z)=(X,X_{\bar{z}})=0
\eeq
and 
\beq
(X_z,X_{zz})=(X_z,X_{z\bar{z}})=(X_{\bar{z}},X_{\bar{z}\bar{z}})=(X_{\bar{z}},X_{z\bar{z}})=0
\eeq
respectively. Define 
\beq
(X_z,X_{\bar{z}}):=2e^{\alpha}\, , \hsp  (X_{z\bar{z}},N):=2H^he^{\alpha} \, , \hsp (X_{zz},N):=A^h\, ,
\label{scndderiv1}
\eeq
where $A^h$ is the Hopf differential and $H^h$ is the mean curvature of the surface described by the solution to the string equation of motion. Since we are concerned here with a minimal area surface (described by the string equations of motion) we will have a vanishing mean curvature and the second equation in (\ref{scndderiv1}) is equal to zero. We want to study what happens when the basis  undergoes a small motion with the hope that second derivative quantities will tell us something useful about the sigma model. For this we  write second derivatives as a linear combination of the basic vectors. Thus we write  
\beq
X_{z\bar{z}}=aX+bX_z+cX_{\bar{z}}+dN\, .
\eeq
Then taking inner product with $X$ gives
$$-a=(X,X_{z\bar{z}})=-(X_z,X_{\bar{z}})$$
which implies $a=2e^\alpha$ due to (\ref{scndderiv1}). Doing same for the other vectors in the basis gives 
\beqa
\nonumber
0=(X_z,X_{z\bar{z}})=c(X_z,X_{\bar{z}}) \hsp  \Longrightarrow c=0  \\ \vsp\nonumber
0=(X_{\bar{z}},X_{z\bar{z}})=b(X_{\bar{z}},X_z) \hsp \Longrightarrow b=0\\ 
0=(N,X_{z\bar{z}})=-d \nonumber
\eeqa
This shows that 
\beq
X_{z\bar{z}}=2e^\alpha X\, ,
\label{eomX2}
\eeq
 which is exactly the equation of motion (\ref{eomX}) taking $\L=2e^\a$. Note that without the minimal area condition, i.e. $H^h=0$, equation (\ref{eomX2}) will have an additional term proportional to $H^hN$. 
Repeating this for the other second derivatives we get \\
$$X_{zz}=\alpha_z X_z-A^h N$$ and  $$X_{\bar{z}\bar{z}}=\alpha_{\bar{z}} X_{\bar{z}}-\bar{A}^h N\, .$$

Consider motion of the basic vectors given by 
\beq
t_{i,z}=U_{ij}t_j, \hspace{0.8in} t_{i,\bar{z}}=V_{ij}t_j\, .
\label{vbasis}
\eeq
Note that since second derivatives are expressed entirely in terms of first derivatives, quantities such as $t_{i,z}$ and $t_{i,zz}$ are all expressible in terms of first derivatives. Consequently, the matrices $U$ and $V$ contain $\alpha$ and its derivatives, and $A^h$ (and its complex conjugate). 
Taking second derivatives and imposing the compatibility condition $t_{i,z\bar{z}} =t_{i,\bar{z}z}$ leads to 
\beq
(U_{ij,\bar{z}}-V_{ij,z})t_j+(U_{ij}V_{jk}-V_{ij}U_{jk})t_k=0\,.
\eeq 
This  equation  may be written in matrix form, after dropping the $t$, as 
\beq
U_{\bar{z}}-V_z+[U,V]=0
\label{cpty}
\eeq
with $U$ and $V$ determined from (\ref{vbasis}) as 

\[ U=\left( \begin{array}{cccc}
0 & 1 & 0 & 0 \\
0 & \alpha_z & 0 & A^h \\
2e^{\alpha} & 0 & 0 & 0\\
0 & 0 & -\frac{1}{2}A^he^{-\alpha} & 0\end{array} \right) , \hspace{0.7in}
V=\left( \begin{array}{cccc}
0 & 0 & 1 & 0 \\
2e^\alpha &0 & 0 & 0 \\
0 & 0 & \alpha_{\bar{z}} & \bar{A}^h\\
0 & -\frac{1}{2}\bar{A}^he^{-\alpha} & 0  & 0 \end{array} \right) \,. \] 
The compatibility equation implies that  $A^h$ is a holomorphic function. Furthermore, it leads to a generalized cosh-gordon equation
\beq
\alpha_{z\bar{z}}-2e^\alpha-\frac{1}{2}A^h\bar{A}^he^{-\alpha}=0 \,.
\label{chgordon2}
\eeq
If in (\ref{chgordon2}) we scale the Hopf differential $A^h \to 2A^h$ and similarly for its conjugate, the  generalized cosh-gordon equation becomes 
\beq
\alpha_{z\bar{z}}-2e^\alpha-2A^h\bar{A}^he^{-\alpha}=0\, .
\eeq
Changing coordinates to $w=\sqrt{A^h}\,z, \bar{w}=\sqrt{\bar{A}^h}\,\bar z$ followed by a transformation of the field $\a \to \tilde \a =\a-\half \log A^h\bar A^h$, the resulting equation is the standard cosh-gordon equation (\ref{chgordon}). This procedure is equivalent to setting $A^h=2$ in (\ref{chgordon2}).  The solution for $\alpha$ in terms of theta functions is found to be 
\beq
\label{alphasol}
\alpha(z,\bar{z}) = 2 \ln \frac{\theta(\zeta(z,\bar{z}))}{\theta\left[\begin{array}{c} \Delta_1/2 \\ \Delta_2/2 \end{array}\right](\zeta(z,\bar{z}))}\, ,
\eeq  
where $\zeta$ is a complex vector of dimension equal to $g$, the genus of the Riemann surface, and $\Delta_{1}$, $\Delta_{2}$ are the {\em characteristics} of the theta function. They are also vectors of integers with dimension $g$.

To connect the scalar field $\alpha$ with the solutions for (\ref{eomX}) we introduce a  hermitian matrix
\beq
\mathbb{X} = \left(\begin{array}{cc} X_0+X_3 & X_1-i X_2 \\ X_1 + i X_2 & X_0-X_3 \end{array} \right) .
\eeq
Then construct Poincar\'{e} coordinates 
\beq
Z = \frac{1}{\mathbb{X}_{22}} , \ \ \ X+iY =\frac{\mathbb{X}_{21}}{\mathbb{X}_{22}}.
\label{PoinX}
\eeq
The final connection is to recognize that (\ref{chgordon2}) is also the compatibility condition for the system of equations \cite{BB}
\beq
\Phi_z=\tilde{U}\Phi \hspace{0.5in} \Phi_{\bar{z}}=\tilde{V}\Phi
\label{lax}
\eeq
where $\tilde{U}$ and $\tilde{V}$ are given by 
\beq
\tilde{U} = \frac{1}{2}\left(\begin{array}{cc} 0 & 2\lambda e^{\alpha/2} \\ 2e^{-\alpha/2} & \alpha_z \end{array} \right) \hspace{0.5in}  \tilde{V} = \frac{1}{2}\left(\begin{array}{cc} \alpha_{\bar{z}} & - e^{-\alpha/2} \\ \frac{2}{\lambda}e^{\alpha/2} & 0 \end{array} \right) . 
\eeq
The parameter $\lambda$ is the spectral parameter which emerges in the study of  integrable differential equations. Given a solution for the pair of equations (\ref{lax}),  the Poincar\'{e} coordinates are related to (\ref{lax}) by 
\beq
X+iY=\frac{\Phi_{11}\bar \Phi_{12} + \bar \Phi_{22} \Phi_{21}}{\Phi_{12} \bar \Phi_{12} + \Phi_{22} \bar \Phi_{22}} \hspace{0.5in} Z=\frac{\sqrt{\det \Phi  \det \Phi^{\dagger}}}{\Phi_{12} \bar \Phi_{12} + \Phi_{22} \bar  \Phi_{22}}
\label{psol}
\eeq
where $\Phi=\left(\begin{array}{cc} \Phi_{11} & \Phi_{12} \\ \Phi_{21} & \Phi_{22} \end{array} \right) $. One advantage of Poincar\'{e} coordinates is that the boundary of \adsn3 is now a copy of $\calr^2$. 

Riemann theta function  solution for the cosh-gordon equation was found in \cite{IKZ, KZ}, then  the system (\ref{lax}) was solved in terms of those theta functions. The solution for $\Phi$ furnishes solutions for  (\ref{psol}) for various values of $|\l| = 1$. In those works the spectral parameter did not coincide with branch points of the auxiliary Riemann surface by which the family of Wilson loops are parametrized. The Wilson loops are in the boundary of Euclidean \adsn3 where $Z = 0$.

%%%%%%%%%%%%%%%%%%%%%%%%%%%%%%%%%%%%%%%%%%%%%%%%%%%%%%%%%%%%%%%%%%%%%%%%%%%%%%%%%%%%%%%%%%%%%%%%%%%%%%%%%%%%%%%%%%%%

\section{New Solutions}
Technically, the major deviation from the previous work \cite{KZ} is that we set one branch point equal to $\l$. We need $|\l| = 1$ and throughout we will use $\l = 1$.  In order to explore the behavior of the solutions under the  restriction that $\l$ and a branch point are equal, we begin by reviewing the previous solutions. 
The solutions for (\ref{psol}) was given in equations (3.94) and (3.95) of \cite{IKZ} as

\beqa
Z &=& \left|\frac{\hat{\theta}(2\int_{p_1}^{p_4})}{\hat{\theta}(\int_{p_1}^{p_4})\theta(\int_{p_1}^{p_4})}\right| 
\frac{|\theta(0)\theta(\zeta)\hat{\theta}(\zeta)|\left|e^{\mu z + \nu \bar{z}}\right|^2}{|\hat{\theta}(\zeta-\int_{p_1}^{p_4})|^2+|\theta(\zeta-\int_{p_1}^{p_4})|^2} \ ,
\label{Zsol}\\
 && \nonumber \\
X+iY &=& e^{2\bar{\mu}\bar{z}+2\bar{\nu}z}\  
  \frac{\theta(\zeta-\int_{p_1}^{p_4})\overline{\theta(\zeta+\int_{p_1}^{p_4})}-\hat{\theta}(\zeta-\int_{p_1}^{p_4})\overline{\hat{\theta}(\zeta+\int_{p_1}^{p_4})}}{|\hat{\theta}(\zeta-\int_{p_1}^{p_4})|^2+|\theta(\zeta-\int_{p_1}^{p_4})|^2}\ , \label{XYsol}
\eeqa
with 
\beq
\mu = -D_{p_3} \ln \theta(\int_{p_1}^{p_4}), \ \ \ \ \nu =  -D_{p_1} \ln \theta\left[\begin{array}{c} \Delta_1/2 \\ \Delta_2/2 \end{array}\right](\int_{p_1}^{p_4}) .
 \label{munudef}
\eeq 
$p_{1}$ and $p_{4}$ are points on the Riemann surface. Concretely, $p_{1} = 0$ and $p_{4} = \lambda$, where the value of  $\lambda$ is taken to lie on the unit circle. 

When, as in the present case, $\lambda$ coincides with a branch point, $Z$ vanishes and $X+iY$ may be singular. The automatic vanishing of $Z$ can occur if $\hat{\theta}(2\int_{0}^{\lambda=\chi})$, where $\chi$ is a branch point equal to one, is automatically zero. This problem can be traced back to the fact that at $\lambda = \chi$ the columns of $\Phi$ fail to be unequal and the system (\ref{lax}) becomes singular in the sense that $\det \Phi=0$.

In order to fix this we must regularize $\Phi$ and choose a new matrix consisting of a column from the old $\Phi$ and a column from the regularized $\Phi_{reg}$. 

The solution for the system (\ref{lax}) is 
{\large
\beqa
\Phi=\left(\begin{array}{cc} v \,\frac{\theta(\zeta+A(\lambda))\,\theta(d)}{\hat{\theta}(\zeta)\,\theta(d+A(\lambda))} \, e^{\psi(\l,z,\bar{z})} &
 v \,\frac{\theta(\zeta - A(\lambda))\,\theta(d)}{\hat{\theta}(\zeta)\,\theta(d - A(\lambda))} \, e^{-\psi(\l,z,\bar{z})} \vspace{0.1in} 
 \\ \frac{\hat{\theta}(\zeta + A(\lambda))\,\theta(d)}{\hat{\theta}(\zeta)\,\theta(d+A(\lambda))} \, e^{\psi(\l,z,\bar{z})} &
 - \,\frac{\hat{\theta}(\zeta - A(\lambda))\,\theta(d)}{\hat{\theta}(\zeta)\,\theta(d - A(\lambda))} \, e^{-\psi(\l,z,\bar{z})} \end{array} \right) \, .
\label{laxsol}
\eeqa
}
$v$ is a function on the Riemann surface equal to $\sqrt{\l}$. $A(\l) = \int_{\infty}^{\l}$, is the Abel map and $d = i(c, \bar{c}, c^{0})^{T}$ with $c\in \mathbb{C}^{n}, \, c^{0} \in \mathbb{R}$ is an arbitrary column  vector.  $\hat{\theta}$ is a shorthand notation  for a theta function with characteristics $\Delta_{1}, \, \Delta_{2}$, and $\psi(\l,z, \bar{z}) = z\, \int_{\infty}^{\l}d\Omega_{\infty} +  \bar z \,\int_{\infty}^{\l}d\Omega_{0}$ where $d\Omega_{\infty, 0}$ are the {\em Abelian differentials of the second kind} . They simplify to (\ref{munudef}) when $\chi \ne \l = 1$.

We introduce a local parameter $q=\sqrt{\l-\chi}$ in the vicinity of $\chi$ and take derivatives $\partial_{q}|_{q=0}$ of (\ref{lax}) . Since $\partial_{q}$ commutes with $\partial_{z, \bar{z}}$ we get 
\beqa
\partial_{z}\partial_{q}\Phi|_{q=0} & = & \partial_{q}\tilde{U}|_{q=0} \, \Phi + \tilde{U}\, \partial_{q}\Phi|_{q=0} 
\nn
\partial_{\bar z}\partial_{q}\Phi|_{q=0} & = & \partial_{q}\tilde{V}|_{q=0} \, \Phi + \tilde{V}\, \partial_{q}\Phi|_{q=0} 
\eeqa
The matrices $\tilde U$ and $\tilde V$ depend on $\l$ only through $\l$ itself, and since $\partial_{q}\l = \partial_{q}(q^{2} + \chi)$, therefore, $\partial_{q}\tilde U$ and $\partial_{q}\tilde V$ both vanish at $q=0$. So the upshot is that if $\Phi$ satisfy (\ref{lax}) then $\partial_{q}\Phi|_{q=0}$ also does. 

The elements of $\Phi_{reg} := \partial_{q}\Phi$ are \footnote{An overall factor of $\frac{\theta(d)}{\theta(d\pm A(\l))}$ which does not depend on $\zeta$ has been dropped. This amounts to dropping a constant multiplier in the determinant of $\Phi_{reg}$ and therefore does not affect the solutions in any significant way.}:
\beqa
\Phi_{reg,11} & = & \frac{v}{\hat \theta(\zeta)} \{\partial_{q} [ \theta(\zeta + A(\l)) e^{\psi(\l,z,\bar z)} ] - \theta(\zeta + A(\l))\, e^{\psi(\l, z, \bar z)} \, \partial_{q} \ln \theta(d+A(\l)) \} 
\nn
\Phi_{reg,21} & = & \frac{1}{\hat \theta(\zeta)} \{\partial_{q} [\hat \theta(\zeta + A(\l)) e^{\psi(\l,z,\bar z)} ] - \hat \theta(\zeta + A(\l))\, e^{\psi(\l, z, \bar z)} \, \partial_{q} \ln \theta(d+A(\l)) \} 
\nn
\Phi_{reg,12} & = & \frac{v}{\hat \theta(\zeta)} \{\partial_{q} [ \theta(\zeta - A(\l)) e^{-\psi(\l,z,\bar z)} ] - \theta(\zeta - A(\l))\, e^{-\psi(\l, z, \bar z)} \, \partial_{q} \ln \theta(d - A(\l)) \} 
\nn
\Phi_{reg,22} & = & -\frac{1}{\hat \theta(\zeta)} \{\partial_{q} [\hat \theta(\zeta - A(\l)) e^{-\psi(\l,z,\bar z)} ] - \hat \theta(\zeta - A(\l))\, e^{-\psi(\l, z, \bar z)} \, \partial_{q} \ln \theta(d - A(\l)) \} 
\nn
\eeqa
where $\l = q^{2} + \chi$. 

Two solutions for (\ref{lax}) with non zero determinants can now be constructed from $\Phi$ and $\Phi_{reg}$ by taking the following combinations;

\beqa
\Phi_1  =  \left(\begin{array}{cc} \Phi_{reg, 11} & \frac{\theta(d-A(\l))}{\theta(d)} \Phi_{12} \\ \Phi_{reg, 21} & \frac{\theta(d-A(\l))}{\theta(d)} \Phi_{22} \end{array} \right) \hsp \Phi_2  =  \left(\begin{array}{cc} \frac{\theta(d+A(\l))}{\theta(d)} \Phi_{11} & \Phi_{reg, 12} \\ \frac{\theta(d+A(\l))}{\theta(d)} \Phi_{21} & \Phi_{reg, 22} \end{array} \right).
\eeqa

In this paper we focus on the solution $\Phi_1$. The new solutions in terms of  Poincar\'{e} coordinates can be derived using (\ref{psol}) as
\beqa
X + i\,Y &=& e^{2\,\psi(\l, z, \bar z)}\,\{x_1(\l, z, \bar z)\,[\partial_{q}\psi(\l, z, \bar z) - \partial_{q} \ln \theta(A(\l)+d) ] + 
x_2(\l, z, \bar z) \} \nonumber\\
\vspace{0.2in}
Z &=&  \frac{\det \Phi_1}{\frac{|v|^{2}\, \theta(\zeta - A(\l))\, \overline{\theta(\zeta - A(\l))} + \hat \theta(\zeta - A(\l))\, \overline{\hat \theta(\zeta - A(\l))}}{\hat \theta(\zeta)^{2}}}
\label{NewSol}
\eeqa

where the functions $x_1$ and $x_2$ are
{\small
\beqa
x_1(\l,\zeta) = \frac{|v|^{2} \, \overline{\theta(\zeta - A(\l))} \, \theta(\zeta + A(\l)) - \overline{\hat \theta(\zeta - A(\l))}\, \hat \theta(\zeta + A(\l))}{\theta(\zeta-A(\l)) \, \overline{\theta(\zeta - A(\l)) }+ \hat \theta(\zeta - A(\l)) \, \overline{\hat \theta(\zeta - A(\l))}} \nonumber
\eeqa
\beqa
x_2(\l, \zeta) =  \frac{|v|^{2} \, \overline{\theta(\zeta - A(\l))} \, \partial_{q}\theta(\zeta + A(\l)) - \overline{\hat \theta(\zeta - A(\l))}\, \partial_{q}\hat \theta(\zeta + A(\l))}{\theta(\zeta-A(\l)) \, \overline{\theta(\zeta - A(\l)) }+ \hat \theta(\zeta - A(\l)) \, \overline{\hat \theta(\zeta - A(\l))}} .\nn
\eeqa
}
The determinant $\det\Phi_1$ is given by 
\beqa
\det \Phi_1 = \k\, \frac{\theta(\zeta)}{\hat \theta(\zeta)}
\label{detphi}
\eeqa

It may be shown that the constant $\k$ is equal to 
\beqa
\k &=& \frac{-v}{ \theta\left[\begin{array}{c} \Delta_1/2 \\ \Delta_0/2 \end{array}\right](0)\,\, \theta\left[\begin{array}{c} \Delta_0/2 \\ \Delta_2/2 \end{array}\right](0)} \nonumber \\
&* &[ \partial_{q}\theta\left[\begin{array}{c} \Delta_1/2 \\ \Delta_0/2 \end{array}\right](A(\l) \, \theta\left[\begin{array}{c} \Delta_0/2 \\ \Delta_2/2 \end{array}\right]( A(\l)) + \theta\left[\begin{array}{c} \Delta_1/2 \\ \Delta_0/2 \end{array}\right]( A(\l))\, \partial_{q} \theta\left[\begin{array}{c} \Delta_0/2 \\ \Delta_2/2 \end{array}\right]( A(\l)) \nonumber \\
& &-\frac{\partial_{q}\theta(d-A(\l)) \theta(d+A(\l)) + \theta(d-A(\l)) \partial_{q}\theta(d+A(\l))}{\theta(d+A(\l)) \theta(d-A(\l))} ]\, .\nonumber \\
\eeqa
One important thing to note here is that upon replacing $\det \Phi_1$ in (\ref{NewSol}) by (\ref{detphi}), it becomes clear that the zeros of $Z$ correspond to the zeros of $\theta(\zeta)$ and $\hat \theta(\zeta)$. In this paper we focus on the Wilson loops that are generated by the zeros of $\hat \theta(\zeta)$.

 The function $\psi$ is bilinear in $z$ and $\bar z$ and is given by;
\beqa
\psi(\l, z, \bar z) =  D_{1} \ln \frac{\theta(0)}{\theta(\int_{0}^{\l})}\,(-z + \bar z).
\label{Odef}
\eeqa
where $D_{1} \ln \frac{\theta(0)}{\theta(\int_{0}^{\l})}$ is a real constant. 
Thus, we have that $\partial_{q}\psi$, which appears in $\Phi_{reg}$, is given by   
\beqa 
\partial_{q}\psi = \partial_{q} \bar \partial \psi \, (-z + \bar z)\, .
\label{dqphi}
\eeqa
Note that $\int_{\infty}^{\l} = \int_{0}^{\l} - \int_{0}^{\infty}$. With these expressions, namely (\ref{Odef}) and (\ref{dqphi}), it can be checked that $\Phi_1$ ($\Phi_2$) satisfies the pair of equations in (\ref{lax}) and the solutions  (\ref{NewSol}) indeed describe the strings moving in $AdS_{3}$. 

%%%%%%%%%%%%%%%%%%%%%%%%%%%%%%%%%%%%%%%%%%%%%%%%%%%%%%%%%%%%%%%%%%%%%%%%%%%%%%%%%%%%%%%%%%%%%%%%%%%%%%%%%%%%%%%%%%%%%%%%%%%%

\section{ The Auxiliary Riemann Surface}
It is important to setup the Riemann surfaces which parametrize our solutions. In particular we will show that  moving one particular  branch point around changes the separation between two correlated Wilson loops. Please refer to  \cite{BB, BBook, RM, LangeB, ThF, FK, GWHB} for a review of \hrss \,  and how they parametrize our solutions. 

\begin{figure}
  \centering
    \includegraphics[width=0.8\textwidth,height=0.5\textwidth]{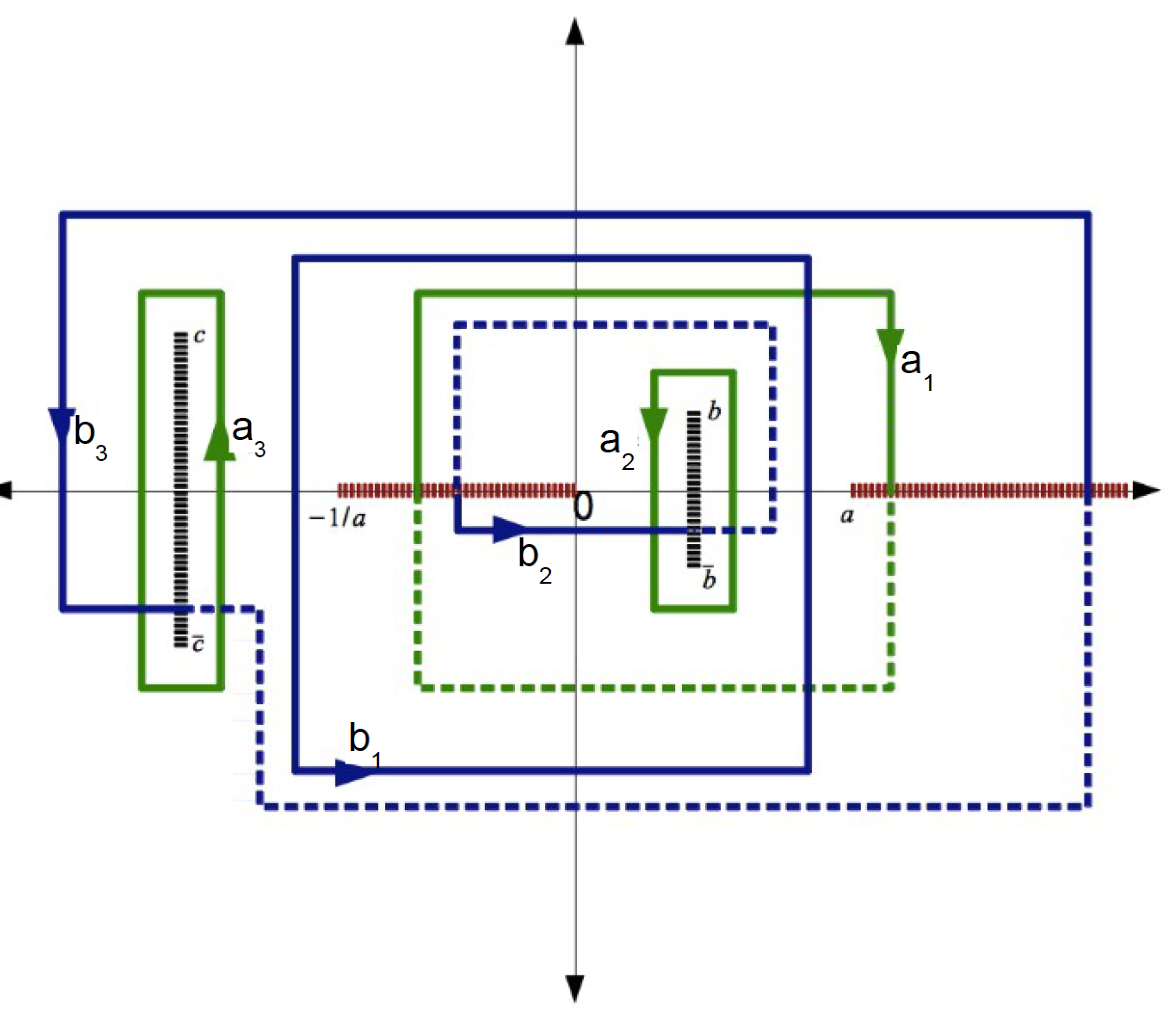}
     \caption{\em g=3 hyperelliptic Riemann surface along with a choice of basis cycles.}
     \label{g3rs}
\end{figure}

Consider a genus  $g=3$ Riemann surface which is given by a hyperelliptic curve  defined by 
\beq
\m^2=\l(\l-a)(\l+1/a)(\l-b)(\l-\bar{b})(\l+c)(\l+\bar{c})\, ,\hspace{0.2in} a\in \mathbb{R}\,, b \ne c \in \mathbb{C}\label{ellipticcurve1}
\eeq
The corresponding \hrs\, with a canonical basis of cycles is displayed in {\em Figure  \ref{g3rs}}. 
Due to the symmetry $\l \to -1/\bar \l$, the branch point $c$ is related to $b$ by $\bar c = -1/\bar b$. This leaves two free 
parameters, $a$ and $b$,  that determine the \hrs. Of these two we  fix $a=\chi=1$ and study when $\l = \chi$. So given $b$  we can compute $\zeta$, and the Riemann period matrix which leads to a solution for $\a$. This solution for $\a$ then leads to  a solution for the string equations of motion. We can change the values of the imaginary or the real part of $b$ and observe the behavior of the Wilson loops that emerge. The real part of $b$ is restricted by $0 < Re(b) < 1$ and the imaginary part is constrained by the condition that $Re(c) = Re(-1/\bar b) < -1$. These restrictions can be deduced by inspection of {\em Figure  \ref{g3rs}}.

\begin{figure}
  \centering
    \includegraphics[width=0.8\textwidth,height=0.4\textwidth]{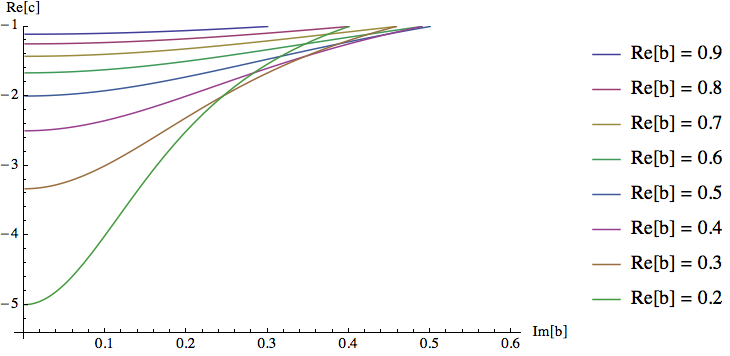}
     \caption{\em Several curves for each value of $Re(b)$ arranged in descending order.}
     \label{bcurves}
\end{figure}

So given a value for $0 < Re(b) < 1$ we can select an appropriate $Im(b)$ and study the solutions we obtain. Since $Re(c)$ is a smooth function of $Im(b)$, we can plot several curves each for a different value of $Re(b)$  as in {\em Figure  \ref{bcurves}}. Each curve describes a set of Wilson loops and as we move along the curve we observe a change in the space of correlated Wilson loops. 

In order to illustrate this, we  study in detail some Wilson loops corresponding to $Re(b)=0.9, 0.6$ and $0.4$.

%%%%%%%%%%%%%%%%%%%%%%%%%%%%%%%%%%%%%%%%%%%%%%%%%%%%%%%%%%%%%%%%%%%%%%%%%%%%%%%%%%%%%%%%%%%%%%%%%%%%%%%%%%%%%%%%%%%%%%%%%%%%

\section{Correlation of Two Wilson Loops}

Let us consider the simpler case of the two Wilson loops separated by a distance $L$. We set $Re(b) = 0.9$ and observe that as we dial the $Im(b)$ from, say, 0.002 to 0.009, the separation $L$ between the two Wilson loops decreases until they start to intersect. In order words one Wilson loop is moving toward the center of the other, and their correlation is expected to become stronger. This may be confirmed by comparing  the computed regularized areas for different values of $Im(b)$ and taking into account that  (\ref{WLcorrelator}) has a negative sign in the exponential function (see {\em Table \ref{resulttab1}}). 

If the characteristic size of the Wilson loops is $a_b$ and their separation is $L$, we are going to consider first the case $L\sim a_b$ and show how the distance $L$ can be changed by changing the positions of the branch point. Afterwards we consider the case where one loop shrinks to a point.

%%%%%%%%%%%%%%%%%%%%%%%%%%%%%%%%%%%%%%%%%%%%%%%%%%%%%%%%%%%%%%%%%%%%%%%%%%%%%%%%%%%%%%%%%%%%%%%%

\subsection{Explicit Results for  Big Loops Correlator with Intermediate $L$}

Let us first work out an example when $L$ is of the order of the size of the Wilson loops. The results are explicit so that the procedure may be easily repeated by any interested person and  for other values of the parameter $b$. 

We can determine the characteristics $\Delta_{1}, \Delta_{2}$ from 
\beqa
 \frac{1}{2}\Omega = \int_{0}^{\infty}\omega = \frac{1}{2}\Delta_{2} + \frac{1}{2}B.\Delta_{1}
\eeqa 
and inspection of {\em Figure  \ref{g3rs}} as
\beqa
\Delta_{1} = \left[\begin{array}{c} 1 \\ 0 \\ 0 \end{array}\right], \hsp \Delta_{2} = \left[\begin{array}{c} 1 \\ 1 \\ 1 \end{array}\right]\, .
\eeqa
The constant vector $d$ may be taken for example to be  $d = i(1+i, 1-i, 1)^{T}$ but for simplicity we use  $d = (0, 0, 0)^{T}$ throughout our calculations. 

The parameter $b$ is chosen from the top most curve in {\em Figure  \ref{bcurves}} corresponding to $Re(b) = 0.9$. We start with   a small value $Im(b) = 0.002$ for the imaginary part of $b$, and compute $\zeta = U\,z + i\, V\, \bar z$  to be;
\beqa
\zeta(z,\bar z) = \left[ \begin{array} {c} 0.3814 \\ 0.0372 \\ 0.1091 \end{array}\right]\, z + 
\left[ \begin{array} {c} 0.3814 \\ 0.1091 \\ 0.0372 \end{array}\right]\, \bar z \, .
\label{zeta}
\eeqa
The  period matrices denoted by $A$ and $B$  are easily computed to be 
\beqa
A = \left(\begin{array}{ccc} 2.4868\, i & -6.1192 & -7.5547 \\
1.2817\, i  & -6.7992 & 6.7992\\
-2.4868\, i & -7.5547 & -6.1192 \end{array} \right) 
\eeqa
\beqa
B = \left(\begin{array}{ccc} i & 0.8250 & -0.8250 \\ 
0.8250 & -0.5+1.8750\, i & 0.0141\, i \\
-0.8250 & 0.0141\, i & 0.5+1.8750\, i \end{array} \right) \, .
\eeqa
Note that as a Riemann period matrix, $B$ is a $3\times3$ symmetric matrix with a positive definite imaginary part.

The matrix 
\beqa
T = - \left(\begin{array}{ccc} 1 & 0 & 0 \\ 0 & 0 & 1 \\ 0 & 1 & 0 \end{array} \right)
\label{Tdef}
\eeqa
allows us to compute the complex conjugate of theta functions as 
\beqa
\overline{\theta\, \left[\begin{array}{c} \a \\ \b \end{array}\right]\, (\zeta)} = \theta\, \left[\begin{array}{c} -T \cdot \a \\ T \cdot \b \end{array}\right]\, (T \cdot \bar \zeta)
\label{thetacomconj}
\eeqa
with $T \cdot B \cdot T = - \overline B$. 
Notice that $T^{2}=1$ and that 
\beq
i\,T \cdot \overline U = V, 
\label{uvtransform}
\eeq
which implies that $ T \cdot \bar \zeta = - \zeta $. Furthermore, we have that $T \cdot \Delta_{1,2} = -\Delta_{1,2}$.  The purpose of this excursion is to say that (\ref{Tdef}), (\ref{thetacomconj}), and (\ref{uvtransform}), along with the evenness of $\theta(\zeta)$ and the oddness of $\hat \theta(\zeta)$ implies that both $\theta(\zeta)$ and $\hat \theta(\zeta)$ are real. Finally it can be checked that $\theta(\frac{1}{2}\Omega)$ is identically zero. 

\begin{figure}
  \centering
    \includegraphics[width=0.5\textwidth,height=0.7\textwidth]{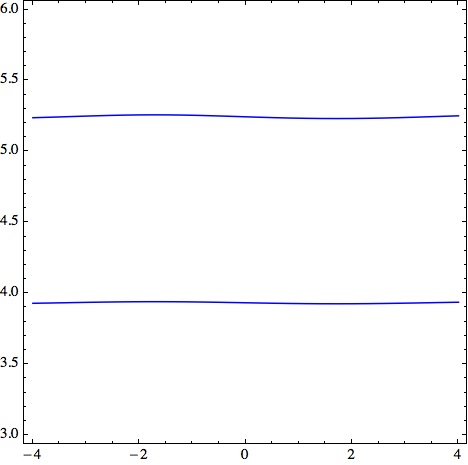}  \hsp
    \includegraphics[width=0.3\textwidth,height=0.7\textwidth]{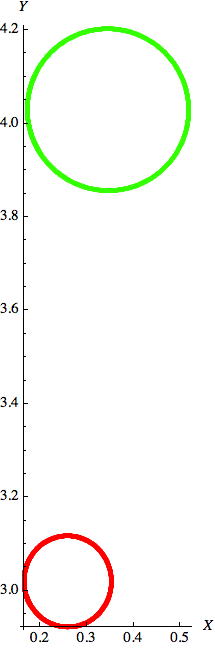}
     \caption{\em Two blue curves describing the zeros of $\hat \theta(\zeta)$ in worldsheet coordinates for $b=0.9+0.002\,i$. The Wilson loops on the right correspond to the pair of blue curves and the region between the curves is the  minimal area surface that connects the two loops.}
     \label{wsnl1}
\end{figure}

\begin{figure}
  \centering
    \includegraphics[width=0.9\textwidth,height=0.6\textwidth]{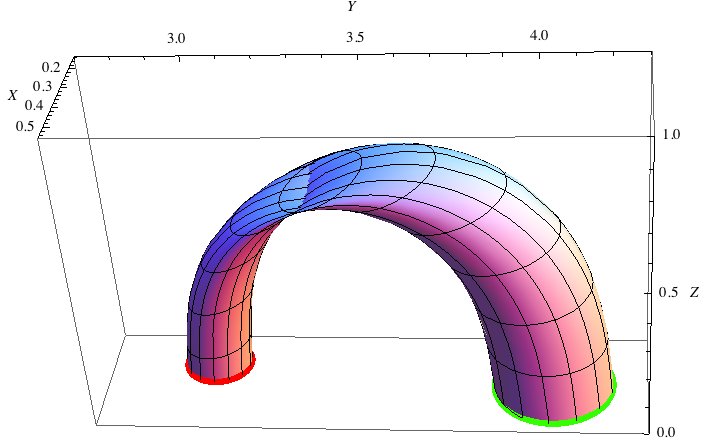}
     \caption{\em  Minimal area surface connecting the loops in  {\em Figure \ref{wsnl1}}.}
     \label{ms1}
\end{figure}

\begin{figure}
  \centering
    \includegraphics[width=0.9\textwidth,height=0.5\textwidth]{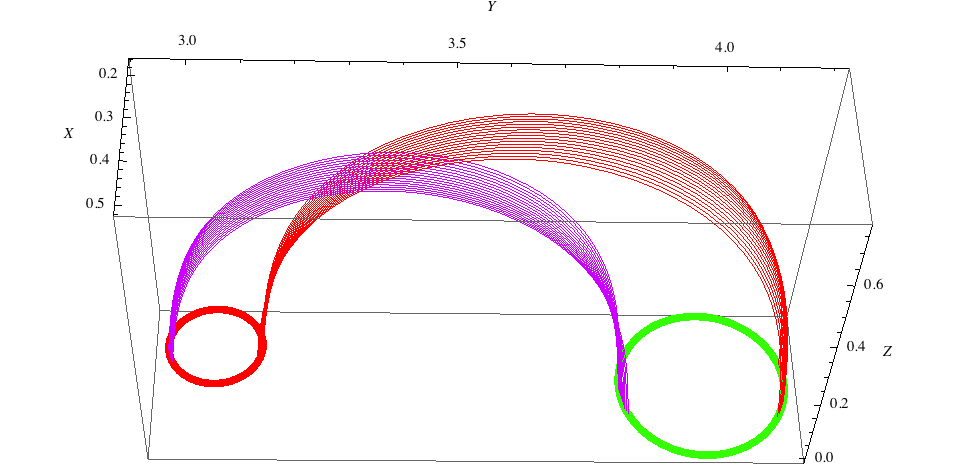}
     \caption{\em  Self intersecting surface connects the loops.}
     \label{ms1b}
\end{figure}

Let's introduce the coordinates of the string worldsheet $z = \s + i\, \tau$, $\bar z = \s - i\, \tau$.  We  plot the curves describing the zero locus of  $\hat \theta(\zeta)$ as in {\em Figure \ref{wsnl1}} . Since the Wilson loops are described by the zeros of $Z$ in the boundary of $EAdS_{3}$,  the pair of open curves therefore correspond to the Wilson loops which are the boundary of the minimal area surfaces in anti-de Sitter space. The region lying between the pair of curves corresponds to the minimal area surface. Using $\tilde X(\zeta) = X(\zeta) + i\, Y(\zeta)$ and $Z(\zeta)$  the curves and the region lying between them are mapped to Euclidean anti - de Sitter space as shown in {\em Figure \ref{ms1}}. From the figure it is clear that one feature of the minimal area surface we find here is that it self intersects.  This is made more explicit in {\em Figure \ref{ms1b}} where lines of constant sigma are depicted.  This feature is present in all the examples we will compute in the rest of this paper. 


Another way to think about these solutions is to take the solution for the concentric Wilson loops in Euclidean anti - de Sitter space \cite{KZ} and slowly move the inner loop to a region outside the bigger loop. The donut-like surface  would deform into something similar to the results here. However as we will show subsequently, many other higher order correlators can be computed here that are not  related to our previous results in this simple way.

%%%%%%%%%%%%%%%%%%%%%%%%%%%%%%%%%%%%%%%%%%%%%%%%%%%%%%%%%%%%%%%%%%%%%%%%%%%%%%%%%

\subsection{Big Loops Correlator with Small $L$}

Now that the procedure for our study of Wilson loop correlator has been made explicitly clear in the example for intermediate $L$, we will present the results for the example when  $b = 0.9 + 0.007\,i$. Remember that the loops approach each other as we increase the imaginary part of $b$ from 0.002 to 0.009. For values well beyond that point we begin to see the emergence of multiple loops from which one can compute the correlator of more than two Wilson loops. This pattern of transition from correlator of two loops to more than two loops is present in all the lines corresponding to the various real part of $b$ plotted in {\em Figure \ref{bcurves}}. However, these transitions occur along the various curves for different values of the imaginary part of $b$.  

One motivation for computing the case of $Im(b) = 0.007$ is to show how the results for the same pair of open curves in {\em Figure \ref{wsnl1}} compares when the $Im(b)$ is increased. We know that as the loops get closer this indicates a stronger correlation, so we expect to see an increase in the  absolute value of the finite part of the area. 

\begin{figure}
  \centering
    \includegraphics[width=0.4\textwidth,height=0.6\textwidth]{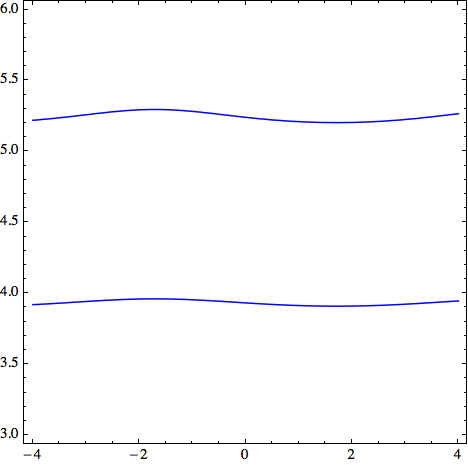}  \hsp
    \includegraphics[width=0.4\textwidth,height=0.6\textwidth]{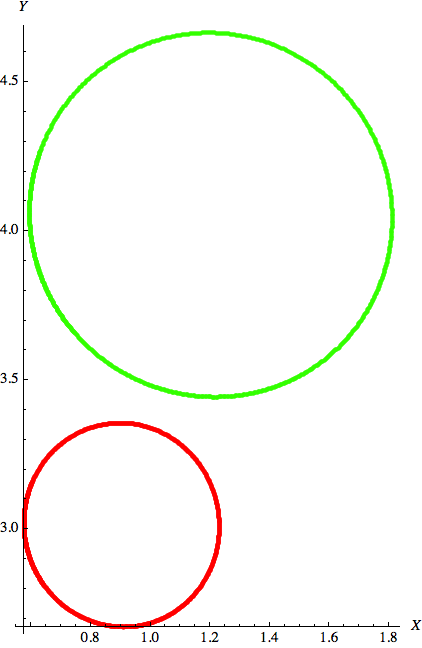}
     \caption{\em Two blue curves describing the zeros of $\hat \theta(\zeta)$ in worldsheet coordinates for $b=0.9+0.007\,i$. The Wilson loops on the right correspond to the pair of blue curves and the region between the curves is the  minimal area surface that connects the two loops.}
 \label{wsnl2}
\end{figure}

\begin{figure}
  \centering
    \includegraphics[width=0.8\textwidth,height=0.5\textwidth]{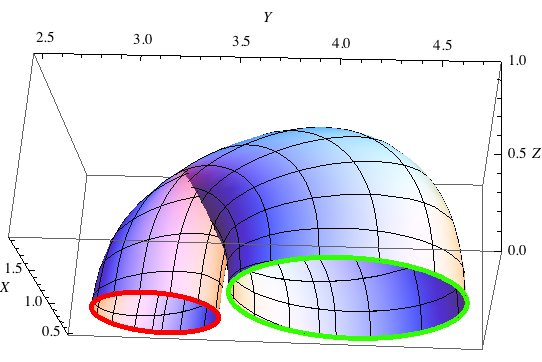}
     \caption{\em  Minimal area surface connecting the loops in  {\em Figure \ref{wsnl2}}.}
     \label{ms2}
\end{figure}

Again the \hrs\, is as in {\em Figure \ref{g3rs}} and this leaves us with the same theta characteristic as before. In this case the corresponding pair of open curves and Wilson loops are shown in  {\em Figure \ref{wsnl2}} and the minimal surface in  {\em Figure \ref{ms2}}.


We have a full analytic result, (\ref{adiv}) and (\ref{eqafinite}), for computing the regularized area for the minimal area surface connecting the  Wilson loops.  We applied Stoke's Theorem in separating the finite part of the area from the divergent part and showed that the divergent part is the length of the Wilson loop (see the {\em Appendix}). Numerically the total area is computed at several different values of $\e$ using the formula $A=4\int d\s\, d\tau e^{\a}$ where $\a= 2 \log \frac{\t}{\hat \t}$. We then fit the data to the linear model $\e \, A= \e \, A_{finite}+L$ and find  both the regularized area and the length of the Wilson loops. In {\em Table \ref{resulttab1}}, we show these results, confirming that the closer the Wilson loops the more strongly correlated they are.

 \begin{table}
 \centering
 \caption{\em Area of connecting  surfaces computed using both numerical and analytical methods}
 \label{resulttab1}
   \begin{tabular}{| c | c | c |}
     \hline
   b &  Area by numerical method  & Area by (\ref{adiv}) and (\ref{eqafinite})  \\ \hline
  $0.9 + 0.007\,i$ &  $ -16.3934+ \frac{19.8020}{\e} $ \rule[-0.5cm]{0pt}{1.2cm} &  $ -16.3956 + \frac{19.8020}{\e} $  \\ \hline
   $0.9 + 0.002\,i$ &  $ -16.3852 + \frac{19.3767}{\e} $ \rule[-0.5cm]{0pt}{1.2cm} &  $ -16.3853 + \frac{19.3767}{\e} $  \\ \hline
    \end{tabular}
\end{table}
      
%%%%%%%%%%%%%%%%%%%%%%%%%%%%%%%%%%%%%%%%%%%%%%%%%%%%%%%%%%%%%%%%%%%%%%%%%%%%%%%

\subsection{Small Loops Correlator}      

\begin{figure}
  \centering
    \includegraphics[width=0.8\textwidth,height=0.4\textwidth]{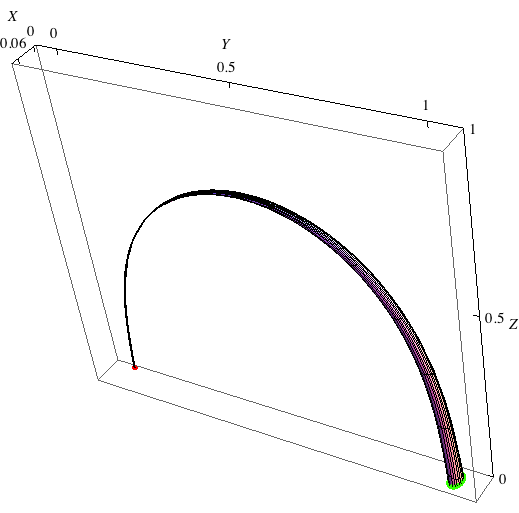}
     \caption{\em  Minimal surface connecting a  small  loop with another loop that has shrunk to a point. $b=0.4+0.002\,i$.}
     \label{ms4}
\end{figure}

We now present the results for the case of small Wilson loops in which one loop shrinks to a point.  An example of such a case is shown in {\em Figure \ref{ms4}} for  $b=0.4+0.002\,i$. If one loop is  very small compared to the other, the small loop may be viewed as a local chiral operator. This operator corresponds to a particle emitted by the worldsheet  that ends on the big loop \cite{BCFM}.  Another possibility is when both loops are small and the distance separating them $L$ is much larger than  their characteristic size, $a_{b}$.  In this situation one can imagine the loops being approximated by local operators with their correlator corresponding to the correlator of two-point functions. This interpretation can also be applied to  cases with more than two Wilson loops.  The area of the surface connecting the small loop to the point is 
\beqa
A = -5.2012 + \frac{6.1328}{\epsilon} \,.
\eeqa

%%%%%%%%%%%%%%%%%%%%%%%%%%%%%%%%%%%%%%%%%%%%%%%%%%%%%%%%%%%%%%%%%%%%%%%%%%%%%%%%%%%%%%%%%%%%%%%%%%%%%%%%%%%%%%%%%%%%%%%%%%%%

\section{Correlator of Three or More Loops}

Now that we have a clear understanding of how  the correlation function for two Wilson loops in the supergravity limit is computed, the idea may  be easily extended to higher number of correlators. Recall that in the two loops case the Wilson loops and their dual minimal surfaces arose from a pair of open curves bounding a horizontal strip in the worldsheet coordinates. It is natural, therefore, to expect that higher number of loops will be obtained from inserting closed curves between a pair of open curves. So for a correlator of three Wilson loops we insert one closed curve, for a correlator of four Wilson loops we insert two closed curves, and so on. The more insertions we add the more complicated the surface becomes. In some instances some of the corresponding Wilson loops may intersect, or may become concentric.  In our previous work \cite{KZ}  all the Wilson loops  obtained were either concentric or, in the case of more than two loops, self-intersecting. The present work is an   extension of those results to  nonintersecting loops. We study an example with three nonintersecting loops. 

\begin{figure}
  \centering
    \includegraphics[width=0.4\textwidth,height=0.6\textwidth]{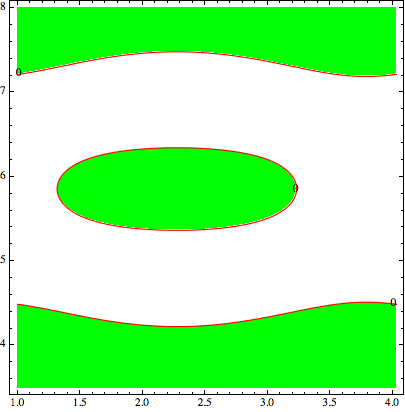}  \hsp
    \includegraphics[width=0.4\textwidth,height=0.6\textwidth]{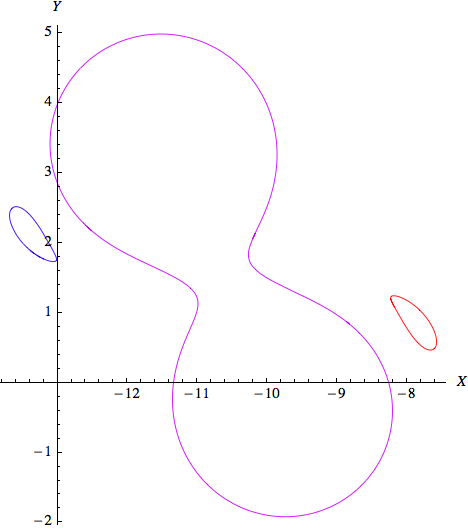}
     \caption{\em Two open curves and a closed curve describing the zeros of $\hat \theta(\zeta)$ in worldsheet coordinates for $b=0.6+0.15\,i$. The small  loops on the right correspond to the pair of open curves and the large  loop to the closed curve.  The region between the pair of open curves and  the closed curve corresponds  to the  minimal area surface that connects the three loops.}
 \label{wsnl3}
\end{figure}

Consider the case when $b=0.6+0.15\,i$ which correspond to the $Re(b) = 0.6$ curve in {\em Figure \ref{bcurves}}. This setup gives correlators of three Wilson loops since the  open curves have a single closed curve in between see {\em Figure \ref{wsnl3}}. In this example we have one large Wilson loop and two relatively small loops. Using the analytic formulae, the area of the surface connecting the three loops is 
\beqa
A = -10.1663 + \frac{26.7055}{\epsilon}\, .
\eeqa

\begin{figure}
  \centering
    \includegraphics[width=0.8\textwidth,height=0.6\textwidth]{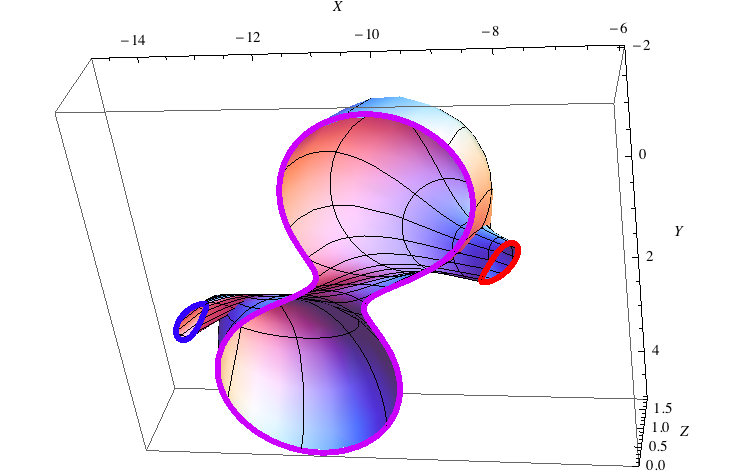}
     \caption{\em  Minimal area surface connecting the three loops in  {\em Figure \ref{wsnl3}}.}
     \label{ms3}
\end{figure}

It might be interesting to consider a limit where the two small loops become point-like.

%%%%%%%%%%%%%%%%%%%%%%%%%%%%%%%%%%%%%%%%%%%%%%%%%%%%%%%%%%%%%%%%%%%%%%%%%%%%%%%%%%%%%%%%%%%%%%%%%%%%%%%%%%%%%%%%%%%%%%%%%%%%

\section{Conclusion}

We have holographically computed examples of correlators of two or more Euclidean Wilson loops. In the case of two loops we show that their separation may be adjusted by changing the location of the branch cut of the auxiliary Riemann surface. In some examples where one or more loops may shrink to a point, such small loops indicate  local chiral operators corresponding to light supergravity modes.  In previous work \cite{KZ} we provided solutions of minimal surfaces ending on several intersecting loops and in view of that the current results may be regarded as an extension  of those results to  nonconcentric and nonintersecting loops of various shapes.

 %%%%%%%%%%%%%%%%%%%%%%%%%%%%%%%%%%%%%%%%%%%%%%%%%%%%%%%%%%%%%%%%%%%%%%%%%%%%%%%%%%%%%%%%%%%%%%%%%%%%%%%%%%%%%%%%%%%%%%%%%%%%

 \section{Acknowledgments}
  
A lot of  debt is owed to M. Kruczenski and  L. Pando Zayas  for the many useful discussions and to M. Kruczenski for hosting S.Z. during which time part of this work was done. 
Also, S.Z. is  grateful to  J. Maldacena for  useful suggestions, and to A. Shapere and S. Das for the many  discussions.  

This work  was supported  by the Lyman T. Johnson Postdoctoral Fellowship.

%%%%%%%%%%%%%%%%%%%%%%%%%%%%%%%%%%%%%%%%%%%%%%%%%%%%%%%%%%%%%%%%%%%%%%%%%%%%%%%%%%%%%%%%%%%%%%%%%%%%%%%%%%%%%%%%%%%%%%%%%%%%

\section{Appendix}

The analytic formulas used in this paper to calculate the area of the minimal surfaces are derived in this appendix.

\subsubsection{Area}

The necessary formulae for computing the regularized area for the minimal area surfaces have already been developed and used in \cite{IKZ, KZ}. We review them here briefly.  In the current context, the area allows us to calculate the correlation of several Wilson loops or the amplitude of exchange of light supergravity modes emitted by the worldsheet ending on a loop. 

Recall that the area of the minimal area surface is given by 
\beq
A=4\int{e^\alpha d\sigma d\tau}\, ,
\label{area2} 
\eeq
which diverges at the boundary due to the vanishing of $\hat \theta(\zeta)$, and therefore requires a regularization \footnote{The integral diverges because $\a = 2 \ln \frac{\theta}{\hat \theta}$, and $\hat \theta$ is zero on the boundary of the surface.}. According to the AdS/CFT prescription one should cut the surface at $Z=\epsilon$ and write the area as
\beq
 A = \frac{L}{\epsilon} + A_{f}\ ,
\eeq
 where $L$ is the total length of the Wilson loops and $A_f$ is the finite part which is identified with the correlator of the Wilson loops
 through:
 \beq
   \langle W^{\dagger}(C_{1} )W(C_{2}) \rangle_{conn}  = e^{-\frac{\sqrt{\lambda}}{2\pi} A_f}\ ,
 \eeq 
where  $\l$ is the 't Hooft coupling of the gauge theory (not to be confused with the spectral parameter). This prescription is equivalent to subtracting the area $A=\frac{L}{\epsilon}$ of a string ending on the contour of length $L$ and stretching along $Z$ from the boundary to the horizon.

Using Fay's trisecant identity, we find an expression for the  exponential in (\ref{area2}) to be a sum of a finite term and a  term that diverges at the boundary where $\hat{\theta}=0$;
\beqa
e^{\alpha} &=& D_{p_1p_3}\ln \theta(0)- D_{p_1p_3}\ln \hat{\theta}(\zeta)\\
           &=&  D_{p_1p_3}\ln \theta(0) - \partial\bar{\partial}\ln \hat{\theta}(\zeta)  . \label{alphaD13}
\eeqa

Integrating the second term at the boundary of the surface obviously leads to divergence so we need to regulate it.
In order to do that we observe we may write $Z$ as a product of a non vanishing function and $\hat{\theta}$
\beq
 Z = |\hat{\theta}(\zeta)| h(z,\bar{z})\ .
\eeq
Substituting (\ref{alphaD13}) into (\ref{area2}) and applying Stoke's theorem we get the expression 
\beq
A = 4 D_{p_1p_3}\ln \theta(0) \int d\sigma d\tau +  \oint \hat{n}\cdot \nabla\ln h \ d\ell
    -  \oint \hat{n}\cdot \nabla \ln Z\ d\ell .
    \label{area3}
\eeq
In \cite{IKZ} we showed that this can  be expressed as 
\beq
A=4D_{p_1p_3}\ln\theta(0)\int{d\sigma d\tau}+\int{d\sigma d\tau \triangledown^2\ln h}-\int{d\sigma d\tau \triangledown^2 \ln Z} \, .
\label{eqn1}
\eeq

The preimage of the  Wilson loops  are a pair of open curves [possibly with closed curves in between] in the world sheet coordinates  as shown in  {\em Figure \ref{wsnl1}}. Therefore,  Stoke's Theorem  is an effective way  to compute the area of the surface bounded by the Wilson loops. This surface corresponds to the horizontal strip between the open curves in the worldsheet coordinates. Hence, integrating along the open curves and along the two vertical boundaries, one at $\s=\s_{0}$ and the other at $\s=\s_{f}$. Here $\s_{0}$ and $\s_{f}$ are two  points along the open curves such that $\s_{f} - \s_{0}$ is the period of $\tilde X(\s, \tau)$, i.e. $\tilde X(\s_{0}, \tau) = \tilde X(\s_{f}, \tau)$.

We parametrize the sine-like curves by the variable $\sigma$, so that the lower curve is now given as $\tau_1(\sigma)$ and the upper one by $\tau_2 (\sigma)$.  According to  Stoke's Theorem, for any smooth real-valued functions $Q$ and $P$ on a regular domain $D$  in $\calr^2$, we have that
\beq
\int_D{\Big(\frac{\partial Q}{\partial x}-\frac{\partial P}{\partial y}\Big) dx dy} = \int_{\partial D}{P dx +Q dy}\, .
\label{eqn2}
\eeq

We apply (\ref{eqn2})  the first term on the right hand side of (\ref{eqn1}) which we denote by $A_{const}$.  Taking $Q=\sigma /2$ and $P=-\tau /2$ we get
 \begin{equation}
A_{const}=-2D_{p_1p_3} \ln \theta(0)\Big(\int_2{\sigma \tau_2'(\sigma)}-\sigma_f\int_3{d\tau}-\int_4{\sigma \tau_1'(\sigma)d\sigma} -\int_2{\tau_2(\sigma)d\sigma}+\int_4{\tau_1(\sigma)d\sigma}\Big)
\label{eqn3}
\end{equation}
where the subscripts 1,2,3,4 on the integrals indicate left, top, right, and bottom boundaries of the domain in the $\sigma-\tau$ plane. Of course  the simplest thing to do here is by following elementary calculus and directly write 
\begin{equation}
A_{const}=4D_{p_1p_3} \ln \theta(0) \int{(\tau_2(\sigma)-\tau_1(\sigma))d\sigma}\\
\label{eqn4}
\end{equation}
However, we chose to arrive at (\ref{eqn3}) by means of (\ref{eqn2}) because it will turn out that this approach is very useful for the other two more complicated terms in (\ref{eqn1}) where the condition that leads to (\ref{eqn4}) is absent.

For the second term in (\ref{eqn1}), denoted by $A_h$, we may take $Q=\partial_{\sigma} \ln h$ and $P=-\partial_{\tau} \ln h$. This gives
\begin{eqnarray}
A_h & = & -\int_1{d\tau\,\partial_{\sigma}\ln h |_{\sigma_0}}-\int_2{d\sigma\,\partial_{\sigma} \ln h |_{\tau_2} \tau_2'(\sigma)}+\int_3{d\tau\, \partial_{\sigma} \ln h|_{\sigma_f}} \nonumber \\
& + & \int_4{d\sigma\, \partial_{\sigma} \ln h|_{\tau_1}\tau_1'(\sigma)}+\int_2{d\sigma\,\partial_{\tau}\ln h|_{\tau_2}}-\int_4{d\sigma\,\partial_{\tau}\ln h|_{\tau_1}}\, .
\label{eqn5}
\end{eqnarray}
Similarly, the last term in (\ref{eqn1}) indicated by $A_z$ becomes,
\begin{eqnarray}
A_z & = & -\Big(\int_1{d\tau\,\partial_{\sigma}\ln Z |_{\sigma_0}}-\int_3{d\tau\, \partial_{\sigma} \ln Z|_{\sigma_f}}\Big)-\int_2{d\sigma\,\partial_{\sigma} \ln Z |_{\tau_2} \tau_2'(\sigma)} \nonumber \\
& + &  \int_4{d\sigma\, \partial_{\sigma}  \ln Z|_{\tau_1}\tau_1'(\sigma)}+\int_2{d\sigma\,\partial_{\tau} \ln Z|_{\tau_2}}-\int_4{d\sigma\,\partial_{\tau} \ln Z|_{\tau_1}}\, .
\label{eqn6}
\end{eqnarray}
Since Z vanishes along the open curves parametrized as $\tau_1(\sigma)$ and $\tau_2(\sigma)$, we see clearly that only the terms in parenthesis in (\ref{eqn6}) are finite leaving all integrals along sides 2 and 4 which are the two horizontal open curves bounding the world sheet to diverge. This is one nice thing about the Stoke's Theorem approach because the divergent part is exposed in very clear manner. To remedy the divergence, we  cut the surface at a height $\epsilon$ very close to the original boundary, and the integrals are no longer divergent up to the boundary of this cut surface.  This is why the string theory is said to have an infrared divergence, but the  corresponding gauge theory has an ultraviolet divergence. The preimage of the boundary of the cut surface is then parametrized by two horizontal curves, $t_1(\sigma), t_2(\sigma)$ that lie very close to the original ones and on the inside of the worldsheet. Once, this is done the formula then becomes 
\begin{eqnarray}
A_z & = & -\Big(\int_{t_1(0)}^{t_2(0)}{d\tau\,\partial_{\sigma}\ln Z |_{\sigma_0}}-\int_{t_1(\sigma_f)}^{t_2(\sigma_f)}{d\tau\, \partial_{\sigma} \ln Z|_{\sigma_f}}\Big)\nonumber \\
& - & \frac{1}{\epsilon}\Big(\int_2{d\sigma\,\partial_{\sigma}  Z |_{t_2}t_2'(\sigma)} - \int_4{d\sigma\, \partial_{\sigma}  Z|_{t_1}t_1'(\sigma)}-\int_2{d\sigma\,\partial_{\tau} Z|_{t_2}} + \int_4{d\sigma\,\partial_{\tau} Z|_{t_1}}\Big) \nonumber \\
\label{eqnaz}
\end{eqnarray}
The first term in parenthesis above consists of integrals along the left and right vertical boundaries of the strip and they vanish. So $A_z=-A_{div}$ where $A_{div}$ is the other grouped item along with its coefficient $1/\epsilon$. Finally, it is clear that the  total area of the minimal area surface can be written as the sum of a term that is finite and a term that diverges as $1/ \e$
\beq
A=A_{conv}+A_{div}
\label{eqnaconv}
\eeq
with
\beq
A_{conv}=A_{const}+A_h\, ,
\label{eqn8}
\eeq
and 
\beq
A_{div}= - \frac{1}{\epsilon}\Big(\int_{\tau_{2}}{d\sigma\,\partial_{\sigma}  Z |_{t_2}t_2'(\sigma)} - \int_{\tau_{1}}{d\sigma\, \partial_{\sigma}  Z|_{t_1}t_1'(\sigma)}-\int_{\tau_{2}}{d\sigma\,\partial_{\tau} Z|_{t_2}} + \int_{\tau_{1}}{d\sigma\,\partial_{\tau} Z|_{t_1}}\Big) \, .\\
\label{adiv}
\eeq

According to the regularization prescription, the divergent part of the total area of the minimal surface should be equal to $L/\e$ where $L$ is the length of the Wilson loop. This implies that if (\ref{adiv}) is correct, it should give  the total length of the Wilson loops.

Recall from  \cite{IKZ}   that the length of the Wilson loop is
\begin{equation}
L=\int_{2+4}{|\triangledown Z|\,dl}-\frac{\epsilon}{2} \int_{2+4} {\frac{\triangledown^2Z}{|\triangledown Z|}\,dl}\, .
\label{eqnlength}
\end{equation}
The regularized area which is the finite part of the total area of the surface ending on the Wilson loops is obtained by 
\beq
A_{finite}=A-\frac{L}{\e}\, .
\eeq
On the other hand we have 
\begin{equation}
A_z=\int_D{\triangledown ^2 \log Z d\sigma d\tau}=\int_{\partial D}{\triangledown \log Z \cdot \vec{dl}}
\end{equation}
where $\vec{dl} =\hat{n}\,dl$ with $\hat{n}$ the outward normal vector. With the tangent vector to the curve given by $(d\sigma, d\tau)$ we take $\vec{dl}=(d\tau, -d\sigma)$.
Going around the loop as before, we obtain 
\begin{equation}
A_z=-\Big(\int_1{d\tau \partial_{\sigma} \log Z |_{\sigma_0} }-\int_3{d\tau \partial_\sigma \log Z |_{\sigma_f}} \Big) 
-\Big(\int_2{\frac{\triangledown Z}{Z}\cdot \vec{dl}}+\int_4{\frac{\triangledown Z}{Z}\cdot \vec{dl}}\Big)
\end{equation}
We have seen that the first term in parenthesis vanishes. Cutting  the surface at $Z=\epsilon$, leaves us with the relation
$$
-A_z=A_{div}=\frac{1}{\epsilon}\int_{2+4}{|\triangledown Z|\,dl}\, , \qquad at \quad Z=\epsilon \, .
$$
Hence, the first term in (\ref{eqnlength}) is exactly equal to the $ \e\,A_{div}$ found in (\ref{adiv}). So the formula for the regularized area of the surface ending on the Wilson loop becomes 
\begin{equation}
A_{finite}=A_{const}+A_h+\frac{1}{2} \int_{2+4} {\frac{\triangledown^2Z}{|\triangledown Z|}\,dl}
\end{equation}
Or more explicitly, 
\begin{equation}
A_{finite}=4D_{p_1p_3} \log \theta(0) \int{(\tau_2(\sigma)-\tau_1(\sigma))d\sigma}+\int_{\partial D}{\triangledown \log h \cdot \vec{dl}}+\frac{1}{2} \int_{2+4} {\frac{\triangledown^2Z}{|\triangledown Z|}\,dl}
\label{eqnafin1}
\end{equation}
From $Z=\hat{\theta} \, h$ we can compute that at $Z=0$ we have 
$\triangledown Z =\triangledown \hat{\theta}\, h $ and $\triangledown^2 Z= \triangledown^2 \hat{\theta}\, h +2 \,\triangledown \hat{\theta} \cdot \triangledown h $. When substituted into (\ref{eqnafin1}) we get
\begin{eqnarray}
A_{finite}=4D_{p_1p_3} \log \theta(0) \int{(\tau_2(\sigma)-\tau_1(\sigma))d\sigma}+\int_{\partial D}{\frac{\triangledown h}{h} \cdot \vec{dl}} \nonumber \\
+\frac{1}{2} \int_{2+4} {\frac{\triangledown^2\hat{\theta}}{|\triangledown \theta|}\,dl}+\int_{2+4}{\frac{\triangledown \hat{\theta}}{|\triangledown \hat{\theta}|} \cdot \frac{\triangledown h}{h}\, dl}
\end{eqnarray}
Looking at the formula for $\triangledown Z =\triangledown \hat{\theta}\, h$ it is clear that $\triangledown Z$ and $\triangledown \hat{\theta}$ are in the same direction so that the unit normal may be taken to be $-\triangledown \hat{\theta}/|\triangledown \hat{\theta}|$. This further simplifies the above equation for $A_{finite}$ giving an expression purely in terms of theta functions and the parametric curves $\tau_1$ and $\tau_2$;
\beqa
 A_{finite}&=&4D_{p_1p_3} \log \theta(0) \int{(\tau_2(\sigma)-\tau_1(\sigma))d\sigma}+\frac{1}{2}\left( \int_{\tau_2(\sigma)}+\int_{\tau_1(\sigma)} \right)
 {\frac{\nabla^2\hat{\theta}}{|\nabla \theta|}\,dl}\,  \nonumber \\
 &=&-2 \Im \left\{ D_{13}\log\theta(0) \oint z d\bar{z}+ \left( \int_{\tau_2(\sigma)}-\int_{\tau_1(\sigma)} \right) D_1 \log\theta(\zeta) d\bar{z}\right\}\,  \nonumber \\
 \label{eqafinite}
\eeqa 
The equivalence of lines one and two of (\ref{eqafinite}) is shown in \cite{KZ}.

%%%%%%%%%%%%%%%%%%%%%%%%%%%%%%%%%%%%%%%%%%%%%%%%%%%%%%%%%%%%%%%%%%%%%%%%%%%%%%%%%%%%%%%%%%%%%%%%%%%%%%%%%%%%%%%%%%%%%%%%%%%%%%%%

\end{document}